\documentclass[twocolumn,final,natbib]{svjour3}
\usepackage[utf8]{inputenc}
\usepackage{graphicx}
\usepackage{amsmath}
\newcommand{\rmd}{{\rm d}}
\newcommand{\rme}{{\rm e}}
\newcommand{\rmi}{{\rm i}}

\begin{document}

\title{Effect of disorder and polarization sequences on two-dimensional spectra of light harvesting complexes
}
\author{Tobias Kramer \and Mirta Rodr\'iguez}
\institute{Tobias Kramer \at Zuse Institute Berlin, Germany
\email{kramer@zib.de} (corresponding author)
\and
Mirta Rodr\'iguez \at Zuse Institute Berlin, Germany
\email{rodriguez@zib.de}
}
\titlerunning{Effect of disorder and polarization sequences on two-dimensional spectra}

\date{10.~October 2019}

\maketitle

\begin{abstract}
Two-dimensional electronic spectra (2DES) provide unique ways to track the energy transfer dynamics in light-harvesting complexes.
The interpretation of the peaks and structures found in experimentally recorded 2DES is often not straightforward, since several processes are imaged simultaneously.
The choice of specific pulse polarization sequences helps to disentangle the sometimes convoluted spectra, but brings along other disturbances.
We show by detailed theoretical calculations how 2DES of the Fenna-Matthews-Olson complex are affected by rotational and conformational disorder of the chromophores.
\end{abstract}

\section{Introduction}

\begin{table*}[bt]
\begin{center}
\begin{tabular}{|c|c|}\hline
$(k,l,m,n)$ & $C_{klmn}$ $\langle 0^\circ ,0^\circ, 0^\circ ,0^\circ \rangle$ \\\hline
$(1, 1, 2, 2)$, $(1, 1, 3, 3)$, $(1, 2, 1, 2)$, $(1, 2, 2, 1)$, $(1, 3, 1, 3)$, $(1, 3, 3, 1)$ & $+\frac{1}{15}$ \\ 
$(2, 1, 1, 2)$, $(2, 1, 2, 1)$, $(2, 2, 1, 1)$, $(2, 2, 3, 3)$, $(2, 3, 2, 3)$, $(2, 3, 3, 2)$ & $+\frac{1}{15}$ \\
$(3, 1, 1, 3)$, $(3, 1, 3, 1)$, $(3, 2, 2, 3)$, $(3, 2, 3, 2)$, $(3, 3, 1, 1)$, $(3, 3, 2, 2)$ & $+\frac{1}{15}$ \\
$(1, 1, 1, 1)$, $(2, 2, 2, 2)$, $(3, 3, 3, 3)$ & $+\frac{1}{5}$ \\\hline
$(k,l,m,n)$ & $C_{klmn}$ $\langle 45^\circ ,-45^\circ, 90^\circ ,0^\circ \rangle$ \\\hline
$(1, 2, 1, 2)$, $(1, 2, 2, 1)$, $(1, 3, 1, 3)$ &  $+\frac{1}{12}$,$-\frac{1}{12}$,$+\frac{1}{12}$ \\
$(1, 3, 3, 1)$, $(2, 1, 1, 2)$, $(2, 1, 2, 1)$ & $-\frac{1}{12}$,$-\frac{1}{12}$,$+\frac{1}{12}$ \\
$(2, 3, 2, 3)$, $(2, 3, 3, 2)$, $(3, 1, 1, 3)$ & $+\frac{1}{12}$,$-\frac{1}{12}$,$-\frac{1}{12}$ \\
$(3, 1, 3, 1)$, $(3, 2, 2, 3)$, $(3, 2, 3, 2)$ & $+\frac{1}{12}$,$-\frac{1}{12}$,$+\frac{1}{12}$\\\hline
\end{tabular}
\end{center}
\caption{$C_{klmn}$ coefficients for isotropic averaging of the $\langle 0^\circ ,0^\circ, 0^\circ ,0^\circ \rangle$ and $\langle 45^\circ ,-45^\circ, 90^\circ ,0^\circ \rangle$ polarization sequences
}\label{tab:Cklmn}
\end{table*}

The investigation of energy transfer in light-harvesting complexes (LHCs) is naturally performed by optical spectroscopy.
To study the time scales and intermediate steps associated with the energy transfer from the antenna to the reaction center requires time- and frequency resolving methods \citep{Blankenship2014}.
One example are transient absorption spectra which probe the response of the system after an excitation (pump) pulse. 
Transient absorption spectra can be described as projected two-dimensional electronic spectra (2DES), which in turn rely on a sequence of four pulses to separate the excitation and emission frequencies.
The theoretical modeling of 2DES within non-perturbative theories of the exciton dynamics requires considerable computational effort due to the possibility of excited state absorption, giving rise to the presence of two excitons in the molecular complex \citep{Mukamel1995,Cho2008,Hamm2011}.
The excitons are coupled to vibrations of the molecules, which induces decoherence and dissipation.
The dissipation directs the excitons to lower lying states, where often the reaction center of LHCs is located \citep{Blankenship2014}.

The first-principle calculation of excitonic site-energies and inter-pigment couplings requires a mixed molecular-dynamics quantum-chemistry approach due to the modulations of the excitonic energies by the solvent and the protein scaffolding, see \cite{Olbrich2011,Aghtar2013}.
For typical distances of chromophores in LHCs and at physiological temperatures, a commonly used model is the Frenkel-type Hamiltonian, which considers the excitons as ``system'' and the vibrational states of the molecule as environment (``bath'') \citep{May2008}.
For LHCs the excitonic couplings, temperature, and reorganization energy can be of comparable magnitude, which precludes the application of weak coupling approximations.
The hierarchical equation of motion (HEOM) method introduced by \cite{Tanimura1989} solves the Frenkel exciton-model dynamics in an exact way and does not rely on small parameter assumption \citep{Ishizaki2009,Kreisbeck2011}.
A comprehensive comparison of Redfield and HEOM spectra for the Fenna-Matthews-Olson (FMO) complex is given by \cite{Hein2012,Kramer2018e}, for the Photosystem~I supercomplex and Foerster theory by \cite{Kramer2018d}, and for combined Foerster-Redfield theory applied to the light harvesting complex II (LHCII) by \cite{Kreisbeck2014a,Novoderezhkin2017}.

Here, we discuss how the protein structures of LHCs are reflected in 2DES for different pulse polarizations and how rotational and static disorder affects the spectra.
This analysis is required to interpret sequences of 2DES in terms of signature of coherent exciton dynamics.

\begin{figure*}[tb]
\begin{center}
\includegraphics[width=0.95\textwidth]{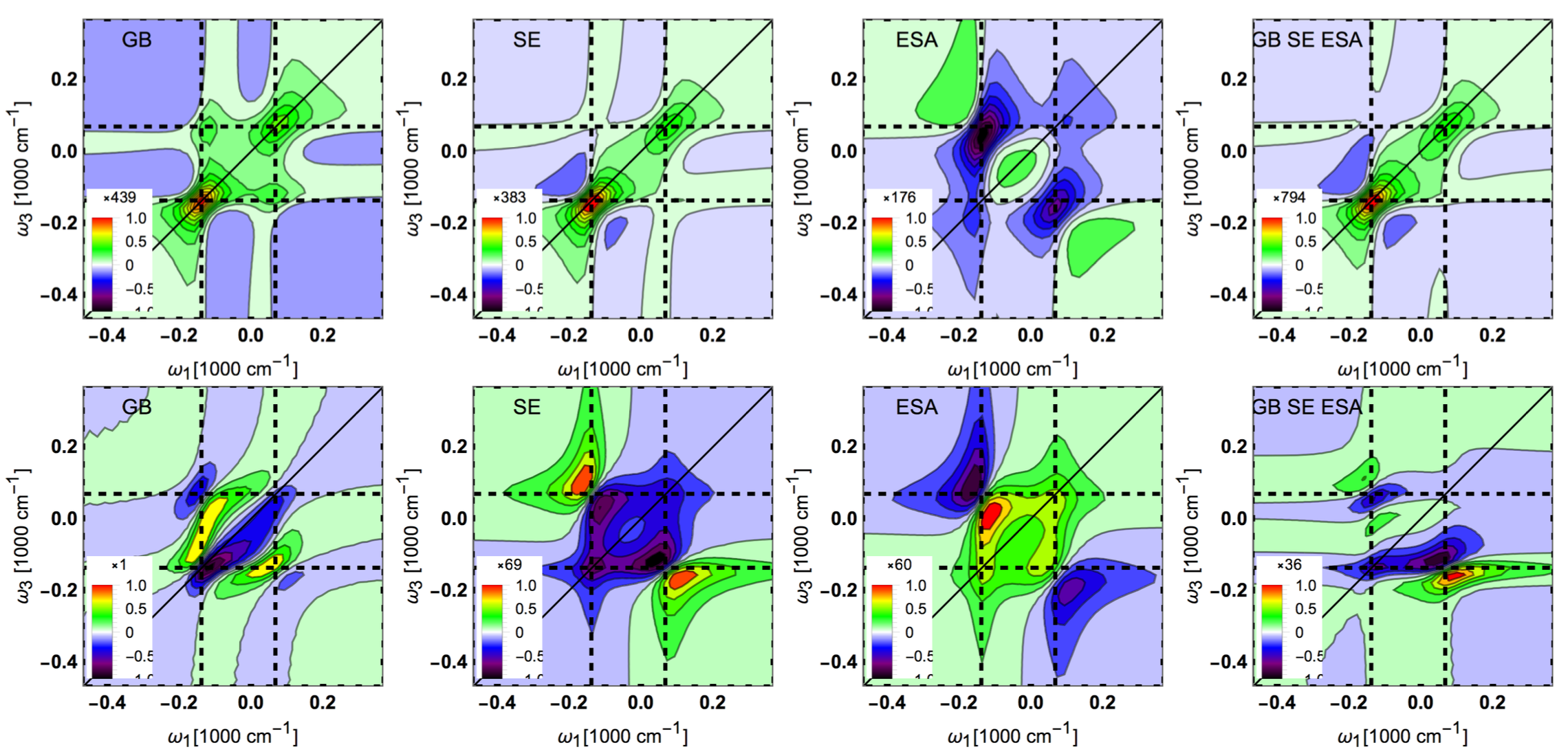}
\end{center}
\caption{Calculated 2DES of a dimer.
Upper row: Real part of the rephasing signal for $\langle 0^\circ ,0^\circ, 0^\circ ,0^\circ \rangle$ polarization sequence at delay time $T_2=40$~fs,
lower row, same for the $\langle 45^\circ ,-45^\circ, 90^\circ ,0^\circ \rangle$ polarizations.
From left to right the contributions of the different pathways are shown, the value above the color bar indicates the magnitude of the signal.
The right panels show the combined, experimentally accessible signal.
In the all-parallel sequence diagonal peaks are enhanced, while in the $\langle 45^\circ ,-45^\circ, 90^\circ ,0^\circ \rangle$ sequence the cross-peaks are visible with additional zero crossings.}
\label{fig:4490DIMER}
\end{figure*}
\section{Two-dimensional electronic spectra (2DES)}

Most commonly 2DES is applied in the ``all parallel'' setup, where all pulses are polarized along the same direction.
In this case, 2DES provides direct insights in to the dynamics of the excitonic energy transfer, since the stimulated emission (SE) and excited state absorption (ESA) signals trace the population dynamics of exciton states during the delay time $T_2$ between the second and third pulse.
For the FMO complex contained in green sulfur bacteria \citep{Olson1962,Fenna1975,Blankenship2014}, the transfer dynamics mirrored in 2DES has been analyzed with HEOM by \cite{Chen2011,Hein2012,Kreisbeck2012}.
In contrast to the SE/ESA contributions, the ground state bleaching (GSB) pathway mirrors the linear absorption signal at long delay times and does not provide direct insights into the population dynamics \citep{Kramer2017a}.
However, GSB is particularly susceptible to vibrational motion, which enters the 2DES \citep{Tiwari2013,Kreisbeck2013} and produces beating signals in the cross-peak amplitudes as function of the delay time $T_2$.
These oscillatory signals have been observed by \cite{Engel2007} in experimental 2DES and need to be separated from electronic coherences.
Over the years various polarization sequences have been explored to disentangle the overlapping and oscillating peaks composing a typical 2DES.

\subsection{Polarization sequences}

The 2DES of light harvesting complexes often shows large and overlapping regions, making it difficult to directly assign peaks to single excitonic energies.
Various extension of 2DES have been proposed, see the reviews by \citep{Nuernberger2015,Lambrev2019} and have been used to isolate specific features in the spectra.
To confirm the structural information of LHCs obtained from crystals \citep{Tronrud2009} and electron microscopy \citep{Bina2016}, circular and linear dichroism studies are important to probe the relative orientations of the transition dipole moments \citep{Lindorfer2018}.
By choosing specific polarization sequences, 2DES is also suitable to interrogate the molecular configuration (orientation and magnitude) of the transition-dipole moments.
To see this, we consider exemplary the stimulated emission rephasing (SE,RP) pathway (see \cite{Kramer2018e}, Eq.~(65)):
\begin{multline}\label{eq:SERPTTT}
S_\text{SE,RP} (T_3,T_2,T_1|p_0,p_1,p_2,p_3)=\\
\rmi{\rm Tr}\big[{\hat\mu}_{p_3}^-(t_3){\hat\mu}_{p_1}^+(t_1)\rho_0{\hat\mu}_{p_0}^-(0){\hat\mu}_{p_2}^+(t_2)\big],
\end{multline}
with time intervals $T_1=t_1$, $T_2=t_2-t_1$, $T_3=t_3-t_2$ and four impulsive interactions of the excitonic states with the light pulses represented by the dipole operators $\mu$ with polarizations $p_i$.
The recorded electric field of the signal is given by the sum of all pathways and incurs an additional conjugation (see \cite{Hamm2011}, Eq.~(4.29)):
\begin{equation}
E_\text{sig} \propto \rmi \left( S_\text{GSB,RP}+S_\text{SE,RP}+S_\text{ESA,RP} \right).
\end{equation}
Within the HEOM method the time-propagation of the coupled exciton-vibrational system is performed numerically.
The first $T_1$ and last $T_3$ interval of the time-dependent signal (\ref{eq:SERPTTT}) are Fourier transformed to the frequency domain and represents the excitation ($\omega_1$)  and emission ($\omega_3$) frequencies.
To account for an ensemble of randomly oriented molecules, a rotational isotropic average of the signal has to be performed.
This leads to a tensorial expression of all possible directional Cartesian components, which is however reduced by symmetry to 21 terms at most \citep{Gordon1968,Yuen-zhou2014,Gelin2017,Kramer2018e}.
Upon transforming the time trace to the frequency domain and isotropic rotational average we obtain
\begin{multline}\label{eq:SERP}
\langle S_{\rm RP}(\omega_3,T_2,\omega_1) \rangle_{\rm rot}
=\int_0^\infty\rmd T_1 \int_0^\infty\rmd T_3
\rme^{-\rmi T_1\omega_1+\rmi T_3\omega_3}\\
\times \!\!\! \sum_{k,l,m,n=1}^{3}  \!\!\! 
C_{klmn} S_\text{SE,RP}(T_3,T_2,T_1|p_{0,k},p_{1,l},p_{2,m},p_{3,n}).
\end{multline}
The $C_{klmn}$ coefficients are determined by
\begin{align*}
C_{klmn}=&\\
\delta_{kl}\delta_{mn}
&\left[(\mathbf{f}_0{}\mathbf{f}_1)(\mathbf{f}_2{}\mathbf{f}_3)-(\mathbf{f}_0{}\mathbf{f}_2)(\mathbf{f}_1{}\mathbf{f}_3)-(\mathbf{f}_0{}\mathbf{f}_3)(\mathbf{f}_1{}\mathbf{f}_2)\right]/30\\
+\delta_{km}\delta_{ln}
&\left[(\mathbf{f}_0{}\mathbf{f}_2)(\mathbf{f}_1{}\mathbf{f}_3)-(\mathbf{f}_0{}\mathbf{f}_1)(\mathbf{f}_2{}\mathbf{f}_3)-(\mathbf{f}_0{}\mathbf{f}_3)(\mathbf{f}_1{}\mathbf{f}_2)\right]/30\\
+\delta_{kn}\delta_{lm}
&\left[(\mathbf{f}_0{}\mathbf{f}_3)(\mathbf{f}_1{}\mathbf{f}_2)-(\mathbf{f}_0{}\mathbf{f}_1)(\mathbf{f}_2{}\mathbf{f}_3)-(\mathbf{f}_0{}\mathbf{f}_2)
(\mathbf{f}_1{}\mathbf{f}_3)\right]/30,
\end{align*}
where $\mathbf{f}_i$ denotes the unit vector of the electric field of the $i$th pulse.
The rotational average can be used to address specific dipole combinations.
To simulate the experimental ensemble requires additionally to account for slower conformational variations of the chromophores (for instance bending and twisting modes), which lead to fluctuations of the site energies, known as static disorder.
Static disorder requires to run simulations for different site energies and leads to a further blurring of the 2DES, albeit with different effects at changing locations in the $\omega_1$-$\omega_3$ frequency-plane.

\subsubsection*{The $\langle 0^\circ ,0^\circ, 0^\circ ,0^\circ \rangle$ polarization sequence}

In the all parallel polarization sequence, GSB and the combined SE+ESA pathways contribute with similar magnitude to the total 2DES signal, see the analysis of the 2DES of \textit{C.~tepidum} by \cite{Kramer2017a}, Fig.~3. 
The isotropic averaging coefficients are listed in Tab.~\ref{tab:Cklmn}.
Typical 2DES of FMO for the $\langle 0^\circ ,0^\circ, 0^\circ ,0^\circ \rangle$ polarization sequence computed with DM-HEOM \citep{Noack2018a,Kramer2018e} are shown in \cite{Kramer2018e}, Fig.~11.
The Hamiltonian and the dipole directions are listed in \cite{Kramer2018e}, Tab.~1 and Eq.~(77), taken from \cite{Adolphs2006}.
The energy transfer is clearly visible in experiments performed by \cite{Brixner2005} and in theoretical computations \citep{Hein2012,Kreisbeck2012} in form of the growing intensities of the lower cross peaks compared to the diagonal peaks.
On top of the energy decay \cite{Engel2007,Panitchayangkoon2010} reported oscillatory amplitudes, which are interpreted as a combination of ground-state bleach induced vibrational modes and electronic coherences.
The electronic coherences are expected to decay on the time-scale of the combined dephasing and relaxation decoherence time \citep{Kreisbeck2012} of the two eigenenergies at the location of cross-peak. 
The slope of the spectral density towards zero frequency determines the pure dephasing time, while the value of the spectral density at the eigenenergies sets the relaxation rate.
Both contribute to the decay time, see \cite{Kreisbeck2012}, SI, and \cite{Kramer2014}, Fig.~8.
Another manifestation of the spectral density $J(\omega)$ of each pigment $m$ in 2DES is the reorganization energy $\lambda_m$:
\begin{equation}
\lambda_{m}=\int_0^\infty \frac{J_{m}(\omega)}{\pi\omega}\rmd\omega.
\end{equation}
\begin{figure}[t]
\centering
\includegraphics[width=0.37\textwidth]{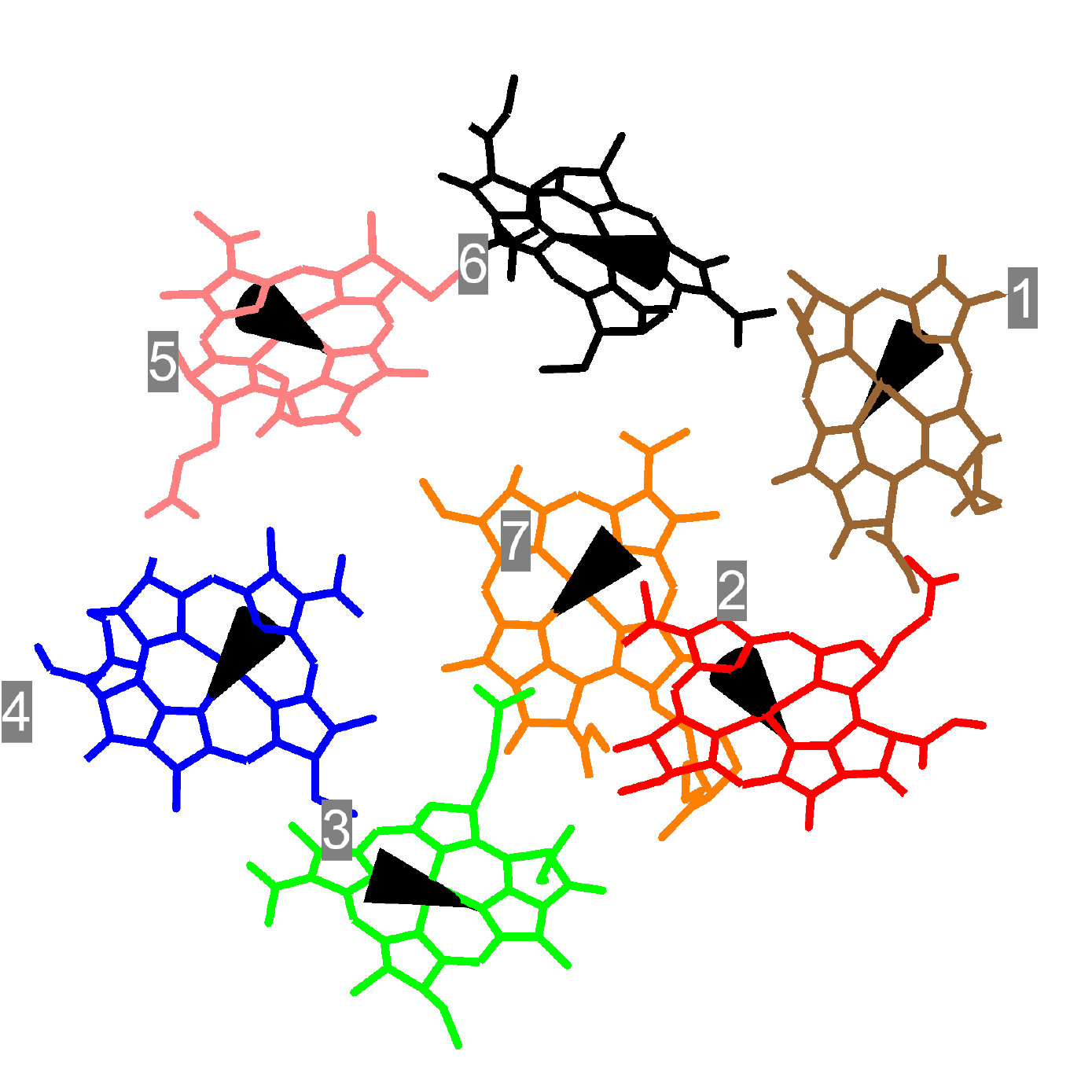}
\caption{The orientations of the transition dipoles in the Fenna-Matthews-Olson complex (FMO monomeric unit 3ENI) along the N$_B$ and N$_D$ nitrogens.}
\label{fig:fmodip}
\end{figure}
The reorganization energy leads to a downward shift of the diagonal and cross peaks with increasing delay \citep{Dostal2016,Kramer2017a}.
The observed shift is consistent with the organization energies around $40$/cm assigned by \citep{Adolphs2006} to the bacteriochlorophylls contained in the FMO complex.

\begin{figure*}[tb]
\begin{center}
\includegraphics[width=0.85\textwidth]{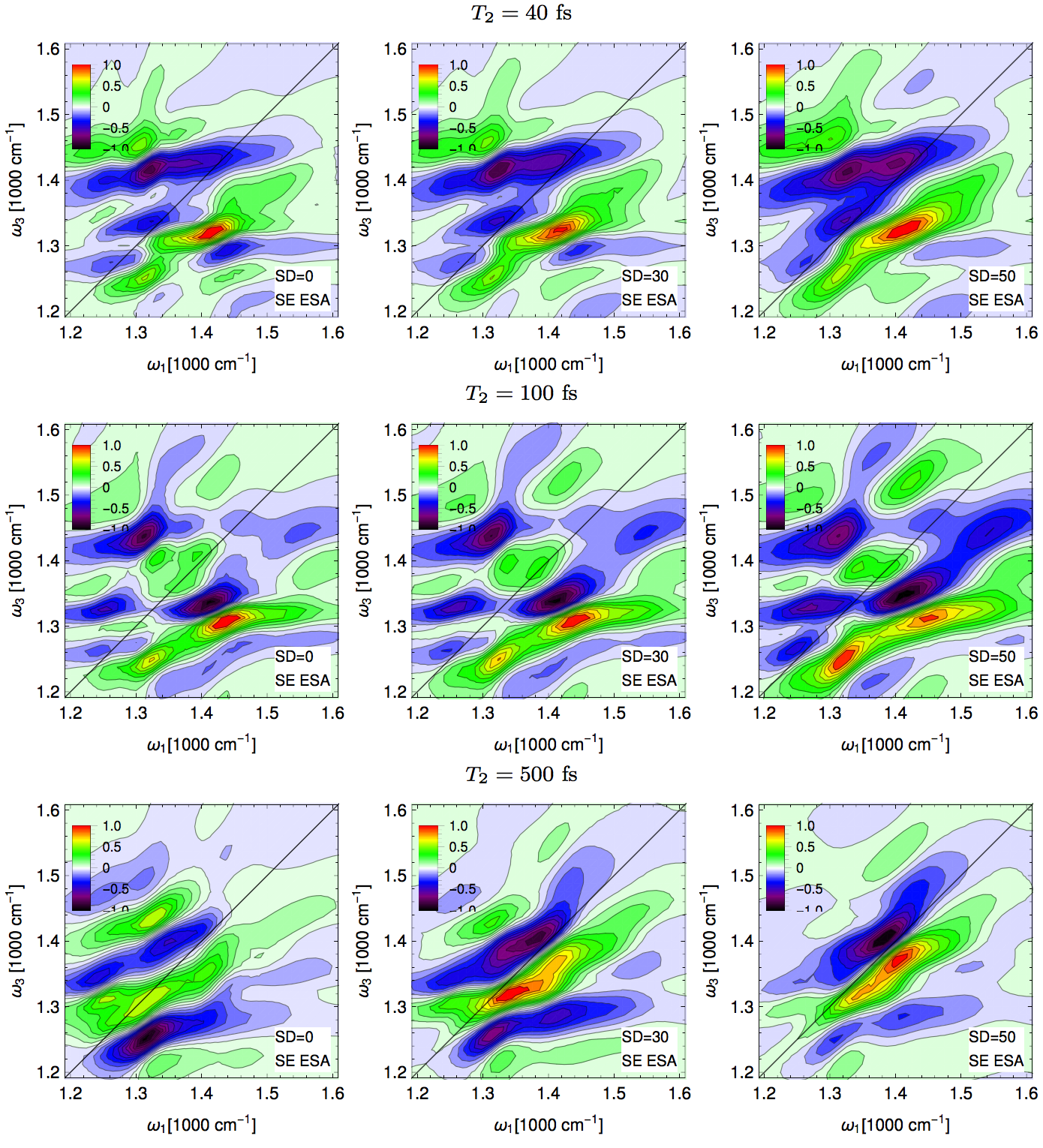}
\end{center}
\caption{Calculated 2DES of FMO for delay times $T_2=\{40,100,500\}$~fs.
Imaginary part of the rephasing signal for $\langle 45^\circ ,-45^\circ, 90^\circ ,0^\circ \rangle$ polarization sequence for increasing values of static disorder (SD), $\{0,30,50\}$~cm$^{-1}$, temperature $T=100$~K. 
See Fig.~5 by \cite{Rodriguez2019b} for the real part of the rephasing signal.
Corresponding measured spectra by \cite{Thyrhaug2018} are shown in their Fig.~2c, SI Fig.~2a, SI Fig.~2b and favor the $30-50$~cm$^{-1}$ static disorder cases.
This results in alternating stripes aligned along the diagonal.}
\label{fig:4490RPDELAYS}
\end{figure*}

\subsubsection*{The $\langle 45^\circ ,-45^\circ, 90^\circ ,0^\circ \rangle$ polarization sequence}

\begin{figure*}[hbt]
\begin{center}
\includegraphics[width=0.85\textwidth]{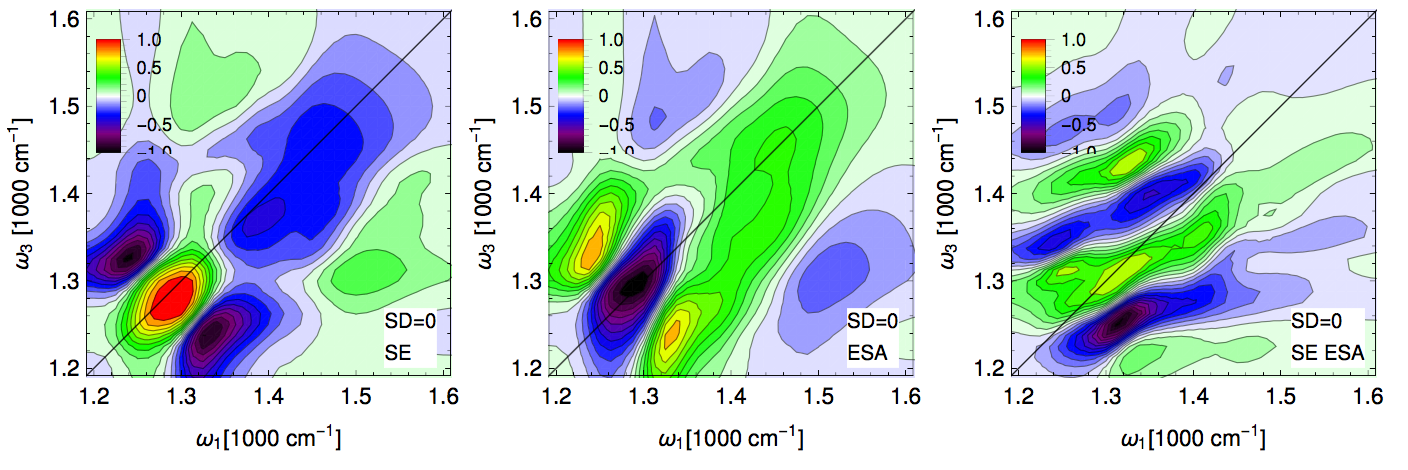}
\end{center}
\caption{Calculated 2DES of FMO. 
Imaginary part of the rephasing signal for the $\langle 45^\circ ,-45^\circ, 90^\circ ,0^\circ \rangle$ polarization sequence at 500~fs delay time in the absence of static disorder. 
Shown is the decomposition of the experimentally accessible signal (sum of SE and ESA, resulting in stripes) into the SE and ESA components (GSB does not contribute).
}
\label{fig:4490RPDELAY500SD1PW}
\end{figure*}

Different polarization directions have been used by\\ \cite{Hochstrasser2001,Zanni2001} to enhance various features in 2D IR spectra  and 2DES \citep{Schlau-cohen2012}.
For a polarization sequence with two orthogonal pulse pairs, the isotropic rotational averaging suppresses the $(1,1,1,1)$, $(2,2,2,2)$, and $(3,3,3,3)$ components, and leads to alternating signs of the remaining $12$ contributions listed in in Tab.~\ref{tab:Cklmn} for the $\langle 45^\circ ,-45^\circ, 90^\circ ,0^\circ \rangle$ case.
This selection of pathways eliminates the ground-state bleaching signal \citep{Westenhoff2012} and emphasizes all dipole pairs with orthogonal orientation.
To demonstrate this, we derive an approximate expression for the 2DES of a dimer with dipole directions given by $\mathbf{D}_1$, $\mathbf{D}_2$ and Hamiltonian 
\begin{eqnarray}
H_\text{dimer}=
\left(
\begin{array}{cc}
-E   &  J \\
J     & +E 
\end{array}
\right)
\end{eqnarray}
with eigenvalues $\{-\Delta E/2,+\Delta E/2=\sqrt{E^2+J^2}\}$.
An analytic expression for the 2DES is obtained by considering a purely coherent dynamics for the density matrix
\begin{equation}\label{eq:rhot}
\rho(t)=\rme^{-\rmi H (t-t_0)/\hbar}\rho(t_0) \rme^{+\rmi H (t-t_0)/\hbar},
\end{equation}
and only introducing decoherence through the inverse of the decoherence time $\alpha=1/\tau_\text{decoherence}$.
Inserting the time-evolution in Eq.~(\ref{eq:rhot}) yields the time-representation of the rephasing signal
\begin{multline}\label{eq:SERP4490t}
\text{SERP}_{4490}(T_1,T_2,T_3)
=2 {(\mathbf{D}_1\times \mathbf{D}_2)}^2 \\
\cos\left( \sqrt{E^2 + J^2} (T_1 + 2 T_2 + T_3)\right)\rme^{-\alpha(T_1+T_2+T_3)}.
\end{multline}
The cross product of the two transition-dipole directions in Eq.~(\ref{eq:SERP4490t}) leads to a vanishing contribution of any parallel dipole component.
The Fourier transform to the frequency domain results in
\begin{multline}
\lefteqn{\text{SERP}_{4490}(\omega_1,T_2,\omega_3)=
{(\mathbf{D}_1\times \mathbf{D}_2)}^2}\\
\times\bigg[\frac{ 
 (2 \omega_1-\Delta E^+) (\Delta E^++2 \omega_3)\rme^{-\rmi {\Delta E} T_2}}{\Gamma}\\
-\frac{(\Delta E^++2 \omega_1) (\Delta E^+-2 \omega_3)\rme^{ \rmi {\Delta E} T_2}
}{\Gamma}\bigg],\\
\Gamma=2 \pi  (\Delta E^--2 \omega_3) (-\Delta E^+-2
   \omega_3)\\
  \times \left(J^2-(E^-+\omega_1) (\omega_1-E^+)\right),
\end{multline}
where we introduced $\Delta E^\pm=\Delta E\pm\rmi2\alpha$,
$E^\pm=E\pm\rmi\alpha$.
At the cross peak location this expression simplifies to
\begin{multline}\label{eq:SERP4490}
\text{SERP}_{4490}(\omega_1=\Delta E/2,T_2,\omega_3=-\Delta E/2)=\\
-\frac{{(\mathbf{D}_1\times \mathbf{D}_2)}^2}{2\pi\alpha^2}
\left(
\rme^{-\rmi  T_2 \Delta E/\hbar}-\frac{\alpha ^2 \rme^{\rmi T_2 \Delta E/\hbar }}
{{(\Delta E+\rmi \alpha )}^2}
\right).
\end{multline}
The ESA rephasing term has the same form (up to an overall sign), albeit with a slightly different decoherence parameter $\alpha'$ due to the coupling to the vibrational states of two separated pigments.
The observed signal is the real part of the difference of the SE and ESA pathways and brings along an additional sign change in the $\omega_1,\omega_3$ plane, see Fig.~\ref{fig:4490DIMER} for a model dimer with energies $E=50$~cm$^{-1}$ and couplings $J=-90$~cm$^{-1}$.
This results in alternating stripes of positive/negative contributions and shifts the highest/lowest intensities away from the cross-peak locations already in the individual pathways (SE,ESA).
This is in contrast to the all-parallel polarization sequence, where ESA has only negative components and SE positive values, which largely stay in place.

The selective excitation of orthogonal dipole orientations leads to a different 2DES of the FMO complex compared to the all-parallel polarization sequence.
In the FMO complex, the transition dipoles are aligned within a few degrees with the nitrogen atoms $N_B$-$N_D$ in the bacteriochlorophylls, see Fig.~\ref{fig:fmodip}.
The dipoles of bacteriochlorophyll pairs $(1, 3)$, $(1, 5)$, $(2, 7)$, $(3, 4)$, $(4, 5)$, $(5, 7)$ are almost orthogonally arranged, but strong excitonic couplings in conjunction to the orthogonal directions are only existing between the neighbouring pigments $(3, 4)$, $(4, 5)$ \citep{Vulto1998,Vulto1999,Adolphs2008}.
The $\langle 45^\circ ,-45^\circ, 90^\circ ,0^\circ \rangle$ sequence thus enhances the contribution of these specific pairs of bacteriochlorophylls in the complex, however at the added complexity of alternating signs in the signal, which are strongly affected by static disorder.
\cite{Thyrhaug2016} measured 2DES with another polarization sequence, taken to be $\langle 90^\circ ,90^\circ, 0^\circ ,0^\circ \rangle$. Corresponding computed 2DES are discussed by \cite{Kramer2018e}, Fig.~12.

While the $\langle 45^\circ ,-45^\circ, 90^\circ ,0^\circ \rangle$ sequence has been suggested and used for studying cross-peak dynamics in FMO, it is more prone to disorder average than the all-parallel configuration.
The alternating signs of the stripes in the real part of the rephasing signal of the $\langle 45^\circ ,-45^\circ, 90^\circ ,0^\circ \rangle$ sequence tend to diminish the signal and only elongated stripes along the diagonal remain.
Numerically, we compute the disorder average from 5000 realizations of disorder added to the site energies with standard deviation $30$~cm$^{-1}$ and $50$~cm$^{-1}$ .
The resulting 2DES are efficiently encoded in a neural network representation following \cite{Rodriguez2019b}.
This encoding allows us to study various disorder realizations, shown in Fig.~\ref{fig:4490RPDELAYS} for delay times $T_2=40$~fs, $100$~fs, and $500$~fs.

The total 2DES is the addition of SE and ESA contributions (see Fig.~\ref{fig:4490RPDELAY500SD1PW}) and develops a stripe-like pattern.
Note, that in the experimental signals recorded by \cite{Thyrhaug2018} Supplementary Fig.~2 (corresponding to the lower two panels in  Fig.~\ref{fig:4490RPDELAYS}) the stripes are almost perfectly parallel to the diagonal, already at shorter delay times (100~fs) compared to the simulation.
A similar theoretical spectrum for the $40$~fs delay time is shown by \cite{Thyrhaug2018}, Fig.~2d, computed with a stochastic averaging approach instead of the tensorial method used here.

\section{Conclusion}

The understanding of 2DES of LHCs requires to ensemble average individual molecular complexes over rotational and static disorder.
This averaging process has been used experimentally in combination with specific polarization sequences with the goal to enhance certain relative pigment orientations (see \cite{Thyrhaug2016,Thyrhaug2018}).
Our calculations of FMO spectra show that the cross-peaks in these sequences are diminished and split by cancellation effects of positive/negative signal contributions from the SE and ESA pathways.
With increasing static disorder elongated structures arise and blur specific cross-peak contributions.
The disorder sensitivity is different compared to the all-parallel 2DES polarization which causes less cancellations due to sign changes in the signal, but is directly affected by broadening of peaks due to static disorder.

The 2DES of the FMO complex demonstrate that the theoretical models of the exciton energy transfer are in general agreement with the experimental observations.
This includes the values for the excitonic couplings, site energies, and vibrational density of states.
The excitonic, vibrational and reorganization energies derived from theory and experiment bring along electronic coherences, typically decaying within 300~fs at ambient temperatures 
\citep{Kreisbeck2012}.
The functional role of these coherences is still debated since \cite{Avery1961}.
\cite{Kramer2014} find that resonances between vibrational states and electronic energy differences can either hinder or enhance transfer times.

For LHCs with more pigments a simultaneous fitting of 2DES, circular and linear dichroism, and transient absorption spectra will be required to further interpret the experimental data and to provide systematic predictions.
A recent review by \cite{Lambrev2019} of the experimental and theoretical 2DES of LHCs comes to a similar conclusion, especially for the light harvesting complex~II (LHC~II).

\paragraph{Conflict of Interest:} The authors declare that they have no conflict of interest.

\begin{acknowledgements}

The work was supported by the German Research Foundation (DFG) grants KR~2889 and RE~1389 (``Realistic Simulations of Photoactive Systems on HPC Clusters with Many-Core Processors'').
We acknowledge compute time allocation by the North-German Supercomputing Alliance (HLRN).
M.R.\ has received funding from the European Union's Horizon 2020 research and innovation programme under the Marie Sklodowska-Curie grant agreement No.~707636. 

\end{acknowledgements}

\end{document}